\documentclass[preprint]{elsarticle}

\usepackage{hyperref,amsmath,amssymb,subfigure}
\usepackage{xspace}
\newcommand{\LCONet}{LCO network\xspace}
\newcommand{\LCOfull}{Las Cumbres Observatory (LCO)\xspace}
\newcommand{\LCO}{LCO\xspace}
\newcommand{\LCONEO}{LCO NEO Follow-up Network\xspace}
\newcommand{\NEOx}{NEOexchange\xspace}
\newcommand{\arcmin}{\hbox{$^\prime$}}
\newcommand{\arcsec}{\mbox{$^{\prime\prime}$}}

\journal{Icarus}

\biboptions{authoryear}

\begin{document}

\begin{frontmatter}

\title{NEOExchange - An online portal for NEO and Solar System science
\tnoteref{mytitlenote}}
\tnotetext[mytitlenote]{Available on the web at \url{https://lco.global/neoexchange}. Code available from GitHub at \url{https://github.com/LCOGT/neoexchange/}}


\author[lco]{T. A. Lister\corref{mycorrespondingauthor}}
\ead{tlister@lco.global}
%orcid: 0000-0002-3818-7769
\cortext[mycorrespondingauthor]{Corresponding author}
\author[lcocardiff]{E. Gomez}
\ead{egomez@lco.global}
\author[lco]{J. Chatelain}
\ead{jchatelain@lco.global}
\author[lco,ucsb,b612,uw]{S. Greenstreet}
\ead{sarah@b612foundation.org}
\author[lcointern,freedomphot]{J. MacFarlane}
\author[lcointern]{A. Tedeschi}
\author[lcointern]{I. Kosic}

\address[lco]{Las Cumbres Observatory, 6740 Cortona Drive Suite 102, Goleta, CA 93117, USA}
\address[lcointern]{Intern at Las Cumbres Observatory, 6740 Cortona Drive Suite 102, Goleta, CA 93117, USA}
\address[lcocardiff]{Las Cumbres Observatory, School of Physics and Astronomy, Cardiff University, Queens Buildings, The Parade, Cardiff CF24 3AA, UK}
\address[b612]{Asteroid Institute, 20 Sunnyside Ave, Suite 427, Mill Valley, CA 94941, USA}
\address[uw]{Department of Astronomy and the DIRAC Institute, University of Washington, 3910 15th Ave NE, Seattle, WA 98195, USA}
\address[ucsb]{University of California, Santa Barbara, Santa Barbara, CA 93106, USA}
\address[freedomphot]{Freedom Photonics LLC, 41 Aero Camino, Santa Barbara, CA 93117, USA}

\begin{abstract}
\LCOfull has deployed a homogeneous telescope network of ten 1-meter telescopes to four locations in the northern and southern hemispheres, with a planned network size of twelve 1-meter telescopes at 6 locations. This network is very versatile and is designed to respond rapidly to target of opportunity events and also to perform long term monitoring of slowly changing astronomical phenomena. The global coverage, available telescope apertures, and flexibility of the \LCONet make it ideal for discovery, follow-up, and characterization of Solar System objects such as asteroids, Kuiper Belt Objects, comets, and especially Near-Earth Objects (NEOs).

We describe the development of the ``\LCONEO{}" which makes use of the \LCONet of robotic telescopes and an online, cloud-based web portal, \NEOx, to perform photometric characterization and spectroscopic classification of NEOs and follow-up astrometry for both confirmed NEOs and unconfirmed NEO candidates. 

The follow-up astrometric, photometric, and spectroscopic characterization efforts are focused on those NEO targets that are due to be observed by the planetary radar facilities and those on the Near-Earth Object Human Space Flight Accessible Targets Study (NHATS) lists. Our astrometric observations allow us to improve target orbits, making radar observations possible for objects with a short arc or large orbital uncertainty, which could be greater than the radar beam width. Astrometric measurements also allow for the detection and measurement of the Yarkovsky effect on NEOs. The photometric and spectroscopic data allows us to determine the light curve shape and amplitude, measure rotation periods, determine the taxonomic classification, and improve the overall characterization of these targets. We are also using a small amount of the \LCONEO time to confirm newly detected NEO candidates produced by the major sky surveys such as ATLAS, Catalina Sky Survey (CSS) and PanSTARRS (PS1).
We will describe the construction of the \NEOx NEO follow-up portal and the development and deployment methodology adopted which allows the software to be packaged and deployed anywhere, including in off-site cloud services. This allows professionals, amateurs, and citizen scientists to plan, schedule and analyze NEO imaging and spectroscopy data using the \LCONet and acts as a coordination hub for the NEO follow-up efforts. We illustrate the capabilities of \NEOx and the \LCONEO with examples of first period determinations for radar-targeted NEOs and its use to plan and execute multi-site photometric and spectroscopic observations of (66391) 1999 KW4, the subject of the most recent planetary defence exercise campaign.
\end{abstract}

\begin{keyword}
Asteroids -- Near-Earth object --  Experimental techniques -- astrometry -- photometry
%https://www.elsevier.com/__data/promis_misc/yicar_key_words_may2013.pdf
\end{keyword}

\end{frontmatter}

\section{Introduction}
\label{sec:intro}

Near Earth Objects (NEOs) are our closest neighbors and research into them is important not only for understanding the Solar System's origin and evolution, but also to understand the consequences of, and to  protect human society from,
potential impacts. NEOs consist of two subclasses, Near Earth Asteroids (NEAs) and a smaller fraction of Near Earth Comets (NECs). NECs are thought to be extinct comets that originally came from the Kuiper Belt or Oort Cloud, whereas NEAs originate in collisions between bodies in the main asteroid belt and have found their way into near-Earth space via complex dynamical interactions. Understanding these interactions, the populations of the source regions, and the resulting orbital element distributions requires accurate orbits for robust samples of the NEO population. Substantial numbers of objects must be observed in order to properly debias the sample and correctly model the NEO population. The existing surveys such as the Asteroid Terrestrial-impact Last Alert System (ATLAS; \citealt{Tonry2018ATLAS}), Catalina Sky Survey (CSS), the PanSTARRS1 (PS1) \& PanSTARRS2 (PS2) surveys \citep{Wainscoat2014} and NEOWISE \citep{Mainzer2014} are not normally capable of following-up their own NEO
candidate detections and require additional programs of NEO follow-up on other telescopes to confirm and characterize the new NEOs.

\LCO has deployed a homogeneous telescope network of ten 1-meter telescopes and ten 0.4-meter telescopes to six locations in the northern and southern hemispheres. These have joined the two existing 2-meter Faulkes telescopes (FTN \& FTS).  The global coverage of this network and the apertures of telescope available make the \LCONet ideal for follow-up and characterization of Solar System objects in general and for Near-Earth Objects (NEOs) in particular.

We describe the creation, including the testing and software development philosophy, of the ``\LCONEO{}" and the central observing portal, \NEOx (often abbreviated as `NEOx'), in Section~\ref{sec:neoexchange}. We illustrate the use of the \LCONEO in Section~\ref{sec:followup} to perform the science cases of NEO candidate follow-up and NEO characterization described above. We summarize some of the results to date and outline plans for future work and development in Section~\ref{sec:results}.

\section{Overview of the \LCO Network}
\label{sec:LCOoverview}

\begin{figure}
\begin{center}
\includegraphics[width=0.8\textwidth]{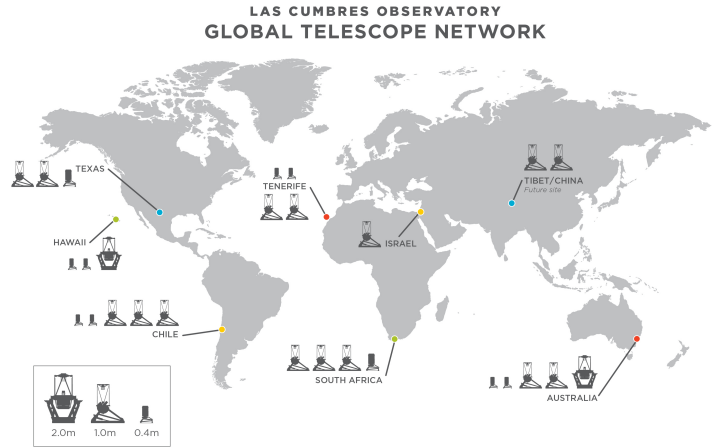}
\end{center}
\caption{Network map of \LCO facilities}
\label{fig:networkmap}
\end{figure}
\LCO completed the first phase of the deployment (see Figure~\ref{fig:networkmap}) with the installation and commissioning of the ten 1-meter telescopes at McDonald Observatory (Texas), Cerro Tololo (Chile), SAAO (South Africa) and Siding Spring Observatory (Australia). These 1-meter telescopes join the two existing 2-meter Faulkes Telescopes which \LCO has operated since 2005. The whole telescope network has been fully operational since 2014 May, and observations are executed remotely and robotically. Two additional 1-meter telescopes for the site in the Canary Islands will be deployed during 2020--2021. Future expansion to a site at Ali Observatory, Tibet is also planned, but the timescale is uncertain and dependent on partner funding.

The 2-meter FTN and FTS telescopes originally had identical $10\arcmin\times10\arcmin$ field of view (FOV) Charge Coupled Device (CCD) imager with 18 filters and a low-resolution ($R\sim 400$, 320 -- 1000\,nm) FLOYDS spectrograph. The imager on the FTN telescope on Maui, HI was replaced in 2020 November with a copy of the MuSCAT2 \citep{Narita2019muscat2} four-channel instrument called MuSCAT3 \citep{Narita2020muscat3}. The 1-meter telescopes have a $26\arcmin\times26\arcmin$ FOV CCD with 21 filters. Each site also has a single high-resolution ($R \sim$\,53,000) NRES echelle spectrograph \citep{Eastman2014NRES, Siverd2018NRES} with a fiber feed from one or more 1-meter telescopes. More details of the telescopes and the network are given in \cite{Brown2013LCOGT}.

For use in the \LCONEO for follow-up and characterization we primarily make use of the 1-meter network, with some brighter targets being observed on the 0.4-meter telescopes. Low resolution spectroscopic observations of NEOs are carried out on the 2-meter FTN and FTS telescopes. With the deployment of the new MuSCAT3 instrument to FTN, which makes use of dichroics to provide simultaneous four-color imaging in $g'r'i'z_s$ filters, we have a powerful tool for simultaneous color and coarse taxonomy determination. This is particularly true for rapid response characterization of small diameter NEOs which would be too faint for spectroscopy.

\section{Overview of the LCO NEO Follow-up Portal: NEOExchange}
\label{sec:neoexchange}

\begin{figure}
\begin{center}
\includegraphics[width=0.95\textwidth]{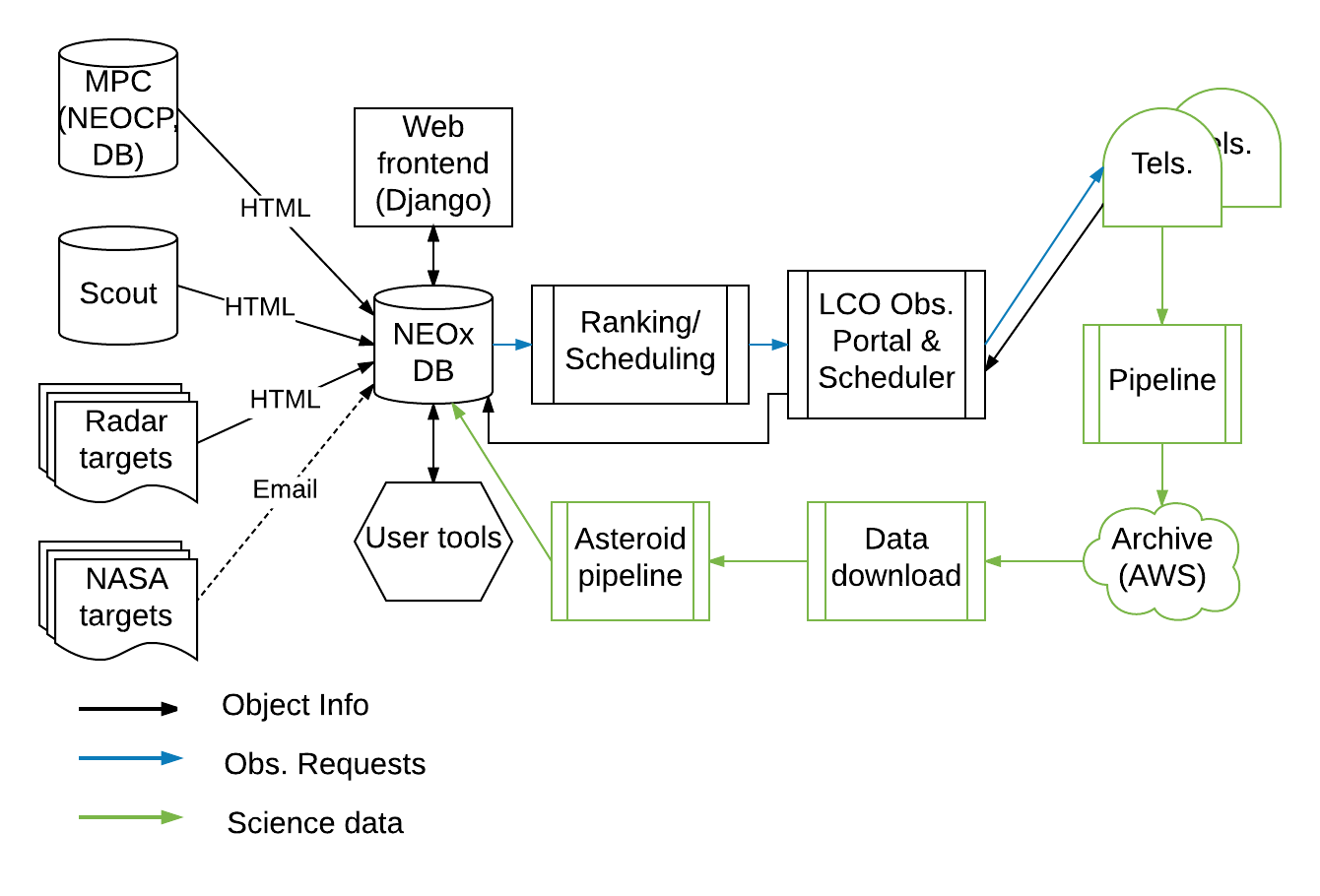}
\end{center}
\caption{Schematic overview of the \NEOx TOM system. Targets for follow-up are ingested into the NEOx DB (center) from the data source (left side), planned and scheduled and sent to follow-up telescopes (right). The resulting data flows through the observatory data reduction pipeline and into the archive. This is retrieved by \NEOx and processed through science specific pipelines and the results of these are also stored in the NEOx DB.}
\label{fig:overview}
\end{figure}

We consider \NEOx to be an example of what are now being called Target and Observation Management (TOM) systems \citep{Street2018TOMs}. The goal of these systems is to ingest a large number of targets of interest and carry out selection of a subset according to some merit function. This subset is then evaluated for observing feasibility and follow-up observations are requested from available telescopes. The resulting status and any data from these observations are returned to the TOM system, which records these details. The results of data reduction on the data are also recorded and this is used as feedback into the next round of observation planning. \NEOx is an implementation of such a TOM system, focusing on Solar System bodies, with a particular emphasis on NEOs.  An outline of the \NEOx system is shown graphically in Figure~\ref{fig:overview}.

The major sections of the \NEOx TOM system are the sources of targets, the database and associated web front-end and interaction tools, the planning and scheduling functions, and the data reduction processes. Each of these will be treated in more detail in the following sections.

\subsection{Data sources and ingest}
\label{sec:ingest}

The major source of target Solar System bodies is the Minor Planet Center (MPC), specifically the MPC's NEO Confirmation Page (NEOCP).\footnote{\url{https://www.minorplanetcenter.net/iau/NEO/toconfirm_tabular.html}} This webpage is parsed to extract the NEO candidates from the surveys that are in need of confirmation. Metadata for each candidate such as the \textsc{digest2} \citep{Keys2019} score (ranging from $0\dots100$, based on how NEO-like the orbit is), arc length, discovery and last observed times are extracted. The orbital elements and observations (in MPC1992 format) are downloaded from the corresponding links. All of these data are ingested into the \NEOx database, updating or creating new records for each of the targets as appropriate.

In addition to the candidates on the NEOCP, we also fetch and parse the list of cross-identifications of objects that were previously on the NEOCP. This allows us to learn the fate of the NEO candidates and update our stored state of the candidates appropriately. This includes when they are designated as a new object, linked with an existing object (either natural or artificial) by the MPC, or the candidate is withdrawn or lost for lack of follow-up or failure to recover it.

Additional sources of targets for characterization efforts (as discussed in more detail in Section~\ref{sec:characterization}) are the planetary radar target lists for Goldstone\footnote{\url{https://echo.jpl.nasa.gov/asteroids/goldstone_asteroid_schedule.html}} and Arecibo\footnote{\url{http://www.naic.edu/~pradar/}}. These are supplemented by potential mission destinations from the Human Space Flight Accessible Targets Study (NHATS) database. The radar targets are obtained by fetching and parsing webpages, whereas the NHATS targets are obtained by parsing emails sent to a mailing list. Both of these methods are somewhat fragile and at risk of missing or misparsing targets if the webpage or email format changes. This illustrates the need for a lightweight protocol that can be used by data requestors to signal targets of interest in a way that is easy and simple to implement as well as parseable and readable by both machines and humans.

Given a list of characterization targets from any of these methods, we then make additional queries in the MPC database in order to extract and store or update the orbital elements of these targets. At present this is also done by parsing webpages (using the \texttt{BeautifulSoup4} library), but the MPC has expressed desires to implement a webservices application programming interface (API) that provides this information in a more robust manner in the future. We will implement and switch to this method of obtaining the information once it becomes available.

Our final source of requests for an increased priority of a particular NEO candidate is from the JPL SCOUT\footnote{\url{https://cneos.jpl.nasa.gov/scout/}} system. This performs analysis of the short-arc NEO candidates that are on the NEOCP and determines if there is any risk of a potential impact or close passage to the Earth. Alerts from the SCOUT system are sent out via email and these can result in the triggering of follow-up, including potentially disruptive `rapid response' that can interrupt already running observations on the \LCO network.

For the NEO candidates from the NEOCP, tasks are periodically executed to update the target lists and cross identifications from the MPC. In the case of the radar and other characterization targets, we update the observations stored in the \NEOx database (DB) and perform orbit fitting to update the orbital elements to the current epoch. The orbit fitting is performed using the \texttt{find\_orb}\footnote{\url{https://www.projectpluto.com/find_orb.htm}} code in a non-interactive mode with the orbit fit to the measurements for the object exported from the database. These tasks are performed with a frequency determined by the typical update frequency of the data source. This is balanced in order to avoid putting undue strain on the remote data source. In the case of the NEO candidates, updates occur every 30 minutes; for the much smaller and less dynamic list of characterization targets, the update tasks are run twice a day.

\subsection{Database, web front-end, and user tools}
\label{sec:db}

\begin{figure}
\begin{center}
\includegraphics[width=0.95\textwidth]{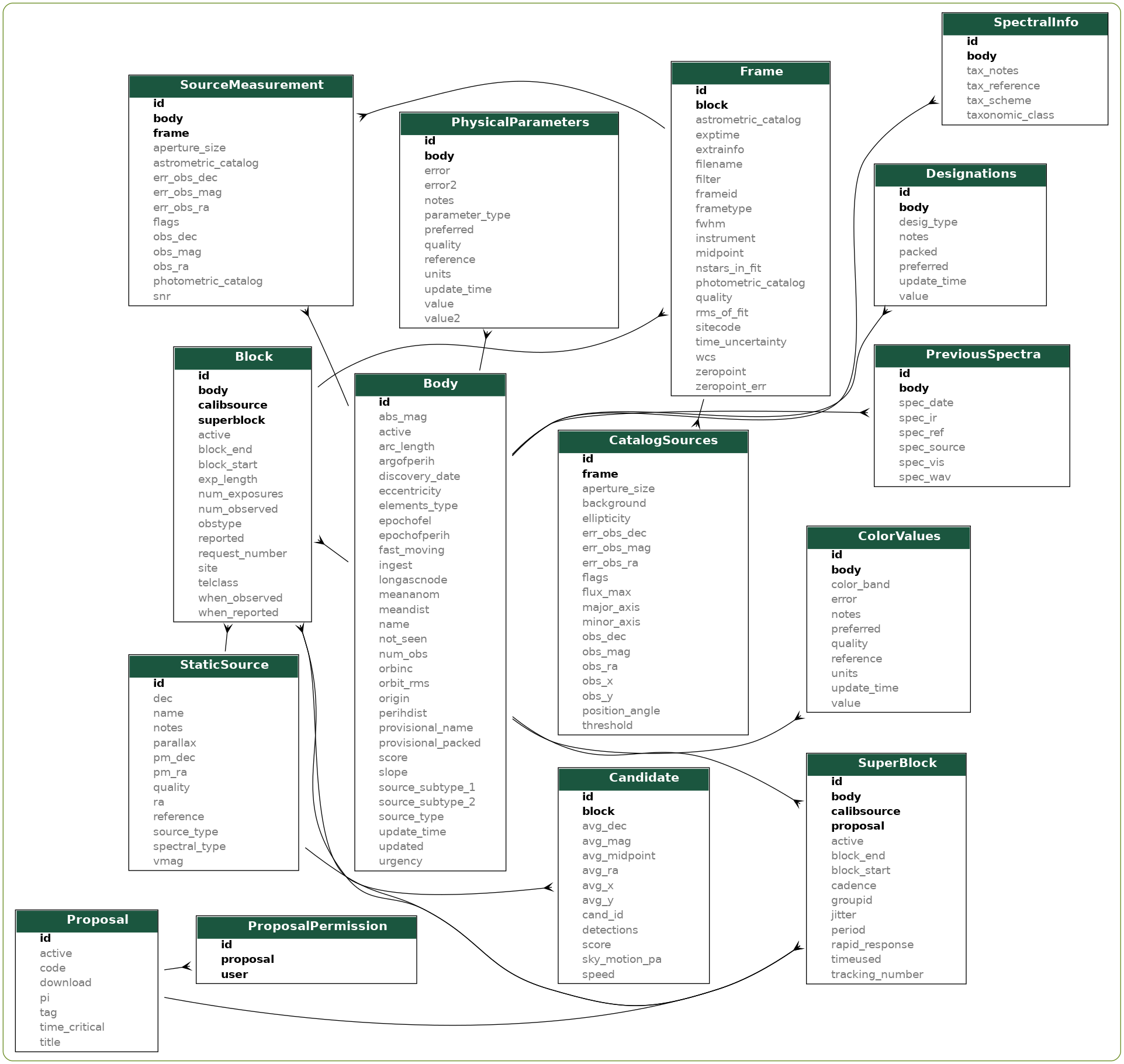}
\end{center}
\caption{Overview of the \NEOx database schema. \textit{Django} Model classes become tables in the database, which are shown as rectangles in the figure with the class/table name on the top. Primary and foreign keys are shown in bold within the tables and the relationships between the tables are shown as black lines between the tables.}
\label{fig:dbschema}
\end{figure}

The targets for follow-up and their associated metadata as described in the previous section, are persisted in a SQL database. This is supplemented by information on user and proposal details for follow-up resources, follow-up requests, data frames obtained, and the catalog products and source measurements derived from those data frames. An overall database schema is shown in Figure~\ref{fig:dbschema}.

We use the \textit{Python} web framework, \textit{Django}\footnote{\url{https://www.djangoproject.com}}, to provide a vendor-independent database abstraction and web front-end. Django provides a means to define a model of a particular object (such as an asteroid target) and the relevant concepts and parameters that are associated with that object, and then have the resulting database table created automatically without the user needing to worry about the low-level database details. It also provides a means to define functions on the model that can provide associated calculations such as a sky position computed from the target model's orbital elements.

As shown in Figure~\ref{fig:dbschema}, the database structure is divided into 3 main areas: 
\begin{itemize}
    \item the \texttt{Body} table and related tables which store additional information (the \texttt{ColorValues, Designations, PhysicalParameters, PreviousSpectra} and \texttt{SpectralInfo} tables). These tables hold the details about the targets (such as target type, the data's origin, orbital elements) and any past color or taxonomic determinations,
    \item the \texttt{SuperBlock} and \texttt{Block} tables that record the details of scheduled observational follow-up, along with the ancillary \texttt{StaticSource} which holds details about static (sidereal) calibration targets,
    \item the \texttt{Frame} table, which stores the metadata about our and others' observations and associated derived quantities resulting from the observed data frames and their data processing. These include the \texttt{CatalogSources} table (the largest in the DB) which holds the source catalog extracted from every frame and the \texttt{SourceMeasurement} table which records definite detections and measurements of a particular \texttt{Body}
\end{itemize}{}
In addition to these major tables, there are a small number of ancillary tables to hold details about the registered users and proposals for the \LCONet and an associated table to link users to proposals, which verifies if they are  allowed to submit observations to the telescopes.

The primary mode of user interaction with \NEOx is through the web front-end. This allows users to view prioritized lists of targets, example details of the targets and the data obtained for them so far. Users can also schedule additional follow-up observations as well as analyze and report data. Certain functions, such as submitting observations, require that the user is registered and authenticated with \NEOx and has been associated with a proposal (or proposals) that have time on the \LCONet.

In addition to the website, various other command-line tools interact with the database. These include those that update the target lists described in the previous section, data download and processing scripts (discussed in Section~\ref{sec:pipeline}), and data analysis tools that can build light curves of targets that have been observed and processed.

\subsection{Planning and scheduling}
\label{sec:planning}

For the purposes of ranking NEO candidates, we compute the following merit function for each candidate:
\begin{equation}
\begin{aligned}
FOM &= \left(e^{\frac{t_\textrm{last}}{t_\textrm{arc}}}-1\right) + \left(e^{\frac{1}{V}}-1\right) + \left(e^{\frac{1}{H}}-1\right)  \\
 &+ 0.5\times e^{\left(\frac{-0.5\times(score-100)^{2}}{10}\right)} + e^{\left(\frac{-0.5\times(SPD-60.)^{2}}{180}\right)}
\label{eqn:FOM}
\end{aligned}
\end{equation}
where $t_\textrm{last}$ is the time since the object was last seen in decimal days, $t_\textrm{arc}$ is the arc length, also in decimal days, $V$ is the current visual magnitude, $score$ is the NEOCP \texttt{digest2} score, $H$ is the absolute magnitude (approx. diameter, assuming an albedo) and $SPD$ is the South Polar Distance in degrees.

This merit function prioritizes targets that have not been seen in a while, have a short arc, have a large diameter (small value of $H$) and are bright, have a high likely NEO `score', and will go directly overhead of our southern hemisphere sites (this last weighting factor is because the \LCONet has many more telescopes in the southern hemisphere; see Figure~\ref{fig:networkmap}).  The first three terms for `last seen'/`arc length', $V$ magnitude, and absolute magnitude ($H$)  in the FOM computation are exponential i.e., for brighter, larger, seen less recently, shorter arc targets, the FOM rises exponentially. The remaining terms involving the `score' and  south polar distance (SPD) are gaussian, where the expected values are 100 and 60 deg, respectively. The `score' term is weighted lower (multiplied by 0.5) than the others to avoid it dominating the priority ranking. The `not seen' and `arc length' parameters are linked together such that targets with high `not seen' values and low `arc length' values (those that haven't been seen in a while and have short arcs) are ranked higher than those with both values high (those that haven't been seen in a while and have longer arcs) or both values low (those that were seen recently and have short arcs).

In order to calculate the values needed for the merit function above and in general, the ephemeris calculations make use of the \textsc{SLALIB} library, which has been wrapped in Python, to handle the position and time calculations. This library provides the basic functions for the Earth, Moon, and object Cartesian positions and velocities, precession/nutation matrices and time conversions for UTC and UT1 to TT and TDB. We have used these routines as the basis for our code that can calculate the brightness, sky position, motion and angle, altitude, and moon separation for a particular site as a function of time. The sky motion is used to set a suitable maximum exposure time that will not produce trailing beyond the size of the seeing disk. The computed ephemeris is then intersected with both the calculated darkness times at the site as well as the telescope class-specific altitude and hour angle limits. The expected brightness of the target is used to set the length of the requested observation block based on 1 magnitude bins. The number of exposures of the previously calculated maximum exposure time that will fit in the block, given the block setup (e.g. slew and settle) and per-frame overheads is calculated. After review by the user, the observation request is sent to the \LCO observing portal and scheduling system for possible scheduling on the \LCONet.{}

For spectroscopy observations, the situation is similar but a little more complicated as additional lamp flat and arc calibration observations are also requested, bracketing the main science spectrum. For calculating the expected signal-to-noise ratio (SNR) on the asteroid target, we make use of a generalized telescope and instrument model to build an Exposure Time Calculator (ETC). This makes use of a parameterized taxonomic class to photometric passband transformation based on \cite{Veres2015} and the mean $(S+C)$ result based on the findings of \cite{Binzel2015} that S- and C-type asteroids account for $\sim50$\,\% and $\sim15$\,\% of all NEOs respectively. We then calculate an effective capture cross-section based on the unobstructed area of the primary mirror. This is then reduced by the throughput, expressed as
\begin{equation}
 t = t_{atm} \times t_{tel} \times t_{inst} \times t_{grating} \times t_{ccd}
\end{equation} where $t_{atm}=10^{(-k/2.5)X}$ is the throughput of the atmosphere, with $k$ being the atmospheric extinction coefficient in the band (in $mag/airmass$) and $X$ is the airmass, $t_{tel}=0.85^{n_{tel\_{mirr}}}$ is the telescope throughput with 0.85 being the typical reflectivity of overcoated aluminium and $n_{tel\_mirr}$ is the number of telescope mirrors, $t_{inst}$ is the instrumental throughput (excluding the grating) which tends to be more uniform with wavelength, $t_{grating}$ and $t_{ccd}$ are the efficiency of the grating and CCD in the observed band respectively. Slit losses are calculated based on the width of the slit and the typical FWHM of the seeing disk.

For the noise sources, we calculate the expected sky background based on the contributions from airglow (we use the 10.7 cm radio flux data as a proxy for the progression through the solar cycle which has been shown to correlate with the airglow intensity e.g. \citealt{Tapping2013solarproxy}), zodiacial light (as a function of ecliptic latitude), the stellar background (as a function of galactic latitude) and the Moon. The model for the brightening due to moonlight is based on \cite{Krisciunas1991}. The readout noise per pixel is also included in the noise sources, though we neglect the dark current as CCD cameras in spectroscopic instruments are normally operated cold enough that the dark current is negligible.

We are developing a more sophisticated and generalized ETC which will make use of an ``ETC language" to describe all of the elements and surfaces (e.g. atmosphere, mirror, lens, grating, CCD etc) and the relationship between them as encountered by a photon from the top of the atmosphere to the detector. For each surface, the ETC language will support use of a scalar (e.g. a target $V$ magnitude), short vector/per-filter values (e.g. extinction per unit airmass as a function of filter/passband) or a ``spectrum" file (e.g. a reflectance spectrum, measured mirror reflectivity as a function of wavelength). This can be combined with the selection of the best-matching atmospheric transmission spectrum from a library of pre-calculated versions (such as the ESO Advanced Sky Model; \citealt{Noll2012}) with selection based on the values of precipitable water vapor, ozone (O$_3$) and aerosols determined from remote sensing. This will be described in more detail in a future publication.

For spectroscopic calibration, nightly flux standards are automatically observed using each FLOYDS instrument. These publicly available flux standards are used to create an airmass curve and account for nightly perturbations in atmospheric transmission. Additionally, for Solar System observations, Solar analogs are required to properly obtain a reflectance spectrum. \NEOx is capable of automatically selecting and scheduling a suitable star from a list of Sun-like options. Stars that are closer on the sky to the target Solar System object are given preference so that observing conditions might be as similar as possible between the analog and the primary target. The exposure time of the analog spectrum is automatically determined based on the brightness of the analog and the resulting spectra are then stored and uploaded to the \NEOx website. Though this process has been fully automated and streamlined, it should be noted that several fundamental limitations are imposed by the \LCO observing portal and scheduler. The greatest of these limitations is the inability to create single observations that include multiple targets. Because of this, we must schedule the target and the analog separately, giving no guarantee that both will be observed on a given night. As neither observation is complete without the other, this adds some risk of lost time to Solar System spectroscopic observations with \LCO. Additionally, due to the constant updating of the \LCO observing schedule, it can be difficult to schedule both target and analog at the same airmass, which is ideal for optimal reduction. We have found, however, that this latter limitation can be mitigated somewhat by use of the aforementioned flux standards so long as both observations were made on the same night.

\subsection{Data reduction pipelines}
\label{sec:pipeline}

During and after the window of validity of the observing requests, we check with the \LCO observing portal whether the request has been executed on the \LCONet. If it has, we check with the \LCO Science Archive for the presence of reduced frames. Data taken on the telescopes of the \LCONet are automatically transferred back to the headquarters in Santa Barbara, CA and pipeline processed. This occurs in near real-time, typically within $\sim10$--$15$\,minutes of shutter close at the telescope. The pipeline assembles any bias, dark, or flat field frames that were obtained at the end of the previous night and start of the current night (and have passed quality control checks) into master calibration frames. If there are no suitable calibration frames from the current night, as can sometimes be the case for flat fields in a specific filter, the most recent master calibration file is retrieved from the calibration library.

The initial data reduction is carried out using the \textsc{BANZAI} pipeline \citep{McCully2018BANZAI} which performs the standard steps of assembling master calibration frames from the individual biases, darks, and flat fields. These calibrations are then applied to the science images to perform bad pixel masking, bias subtraction, dark current correction, and flat field division. Crosstalk correction and gain normalization between the individual quadrants and amplifiers of the Sinistro cameras' Fairchild CCDs are also performed. An astrometric solution is performed using the \textit{astrometry.net} software \citep{Lang2010} which makes use of the 2MASS catalog \citep{2MASS} as input. A catalog of sources detected in the frame (having a certain minimum number of pixels more than $10\sigma$ above the fitted sky background) is also produced using \textsc{SExtractor} \citep{SExtractor1996}. Finally the reduced frames, source catalog, and associated master calibration frames are uploaded to the \LCO Science Archive\footnote{\url{https://archive.lco.global/}} for distribution to end-users. These data products are then retrieved by \NEOx for further astrometric and photometric analysis.

Although the BANZAI pipeline does a good job at removing the instrumental signature from the data and providing a ``first pass" astrometric solution, we cannot make use of this pipeline-produced astrometric solution and source catalog as-is. This is because the camera distortions, although small, are too large for our astrometry goals to be met with a simple linear fit. In addition, because many of our NEO targets are very faint, they will not be included in the relatively shallow pipeline-produced source catalog. To solve both of these problems, we instead re-determine a new astrometric solution, incorporating spatially-varying distortion polynomials. This astrometric solution initially used the PPMXL catalog \citep{PPMXL}, progressed to using the Gaia-DR1 catalog \citep{GaiaDR1} and now makes use of the Gaia-DR2 catalog \citep{GaiaDR2} to derive the astrometric solution and a per-frame photometric zeropoint using the \textsc{SCAMP} software\footnote{\url{https://www.astromatic.net/}}. 

The zeropoint is determined by cross-matching the detected CCD sources with sources in the Gaia-DR2 catalog by position and then iterating, with outlier rejection, to determine the difference in mean magnitude between the instrumental CCD magnitudes and the Gaia catalog $G$ magnitudes. This ignores mismatches between the broad Gaia $G$ passband and the equally broad PanSTARRS-w, which comprise over 85\% of our frames, and does not take into account the source color (almost always unknown) or stellar color. The color correction for reference star colors was not possible when this part of the pipeline was originally written with the absence of accurate colors in either the PPMXL or Gaia DR1 catalogs. Given the large field of view of the LCO telescopes and the corresponding large number (hundreds) of potential reference stars in a typical field, the outlier rejection procedure for the zeropoint is robust against including stars with large discrepancies in color and magnitude. With the availability of $G_{BP}$ and $G_{RP}$ magnitudes in Gaia DR2, and the prospect of spectral types for large numbers of reference stars from low resolution spectra in Gaia DR3, there is the potential to revisit this in the future versions of the pipeline with an improved treatment.

The result of the fitting process and a record of the resulting processed frame data product is stored in the database. After checking that the results of the fits are satisfactory, we perform a source extraction again using the  \textsc{SExtractor} \citep{SExtractor1996} software but to a lower threshold than in BANZAI (3.0 vs 10.0 $\sigma$ above the sky background) to include many more faint sources. The resulting source catalog contents are ingested into the database (forming the largest fraction of the database size). 

This additional processing allows us to extract light curves for any object in the database through either user tools or via the web frontend (see Figure~\ref{fig:overview}). The source catalogs can also be exported to specialized moving object detection software written by the Catalina Sky Survey team \citep{ShellyCSSMOPS}. The results of any detections are also ingested into the database and can be overlaid in a ``Candidate Analyzer" in the web frontend. This allows the user to blink through the acquired frames for that candidate object and the moving object detections are overlaid, along with details about the position (both on the CCD and on-sky), magnitude, and the rate and direction of motion, allowing comparison with the predicted motion to assist confirmation of a NEOCP candidate's recovery. The user can then reject or confirm the identification which will then show a summary of the observation in the MPC1992 80 column format which can then be sent to the MPC. Moving object detections of real but unknown objects can also be confirmed, in which case a new local candidate object is created in the database.

As discussed above, we make use of the Gaia-DR1 \citep{GaiaDR1} and Gaia-DR2 \citep{GaiaDR2} catalogs to perform the astrometric reduction. Gaia-DR1 greatly improved the quality of astrometry obtained by substantially reducing the systematic error contribution by vastly reducing the catalog zonal errors \citep{Spotoetal2017}. The overall astrometric uncertainty is a combination of centroiding error (which is unaffected by the choice of reference catalog),  systematics from the reference catalog and other normally smaller second-order effects such as stellar proper motion and differential chromatic refraction. 

Switching from PPMXL \citep{PPMXL}, which we used prior to the availiability of the Gaia catalogs, to Gaia-DR1/2 has reduced the systematic catalog error from $\sim300$\,mas to $\sim30$\,mas and the overall uncertainty ($\sim0.08\arcsec$--$0.21\arcsec$)
%($\sim0.10\arcsec$--$0.18\arcsec$ and $\sim0.08\arcsec$--$0.21\arcsec$ in the RA and Dec co-ordinates %respectively)
is now dominated by the centroiding error.  With the release of DR2 in April 2018 and the availability of good reference star colors, as well as the release of parallaxes and proper motions in later data releases, it would be possible to take other more subtle effects into account in the astrometric reduction. 

These effects include the differential chromatic refraction (DCR), space motions of the reference stars, and unmodelled optical distortions. DCR is caused by differences in the spectral energy distribution of the Solar System targets and reference stars being refracted differently in the atmosphere, which in terms depends on the variation of temperature, pressure and water vapor \citep{Stone1996} both from site to site and with time. The amount of DCR also depends on the filters used and the optical path length through the atmosphere, which depends on the (changing) zenith distance and hour angle of the target. Historically, the lack of even accurate colors for the majority of reference stars, has made correction of DCR difficult without obtaining large amounts of additional multi-color calibrated photometry. Similarly correcting for the proper motions, and less commonly for the parallax, of the reference stars has been difficult due to lack of accurate available data for the fainter ($V\sim12$--18) stars mostly commonly used as reference stars for $\sim0.5$--2\,m telescopes. Finally, mapping optical distortions that are not modelled by our use of third order distortion polynomials, also requires very accurate reference positions in crowded fields, such as star clusters. With colors, accurate positions and space motions, coming in Gaia DR3 and subsequent releases, these smaller contributions to the astrometric error could be modelled and removed. A more detailed modelling and analysis effort considering the typical targets and observing circumstances of the \LCONEO would be needed to assess the relative contributions of the various sources of astrometric error.

\subsubsection{Spectroscopy Reduction}

The FLOYDS Spectroscopy Pipeline is automatically run on all FLOYDS data obtained by the \LCONet in a manner similar to the BANZAI pipeline described above. The pipeline converts the raw, folded, multi-order fits frames into a 1-dimensional extracted trace of the entire merged spectrum. During this processing, lamp flats are used to minimize fringing in the red branch, then stored flux calibration frames and telluric lines are used for atmospheric correction. The final trace, as well as all intermediate data products, are wrapped in a tarball along with the guider images and made available to users via the \LCO archive. A full description of the data processing, wavelength calibration, and extraction is given on the \LCO website\footnote{\url{https://lco.global/documentation/data/floyds-pipeline/}}. 
The data resulting from this automated pipeline is typically of sufficient quality to determine a rough spectral slope for the target and serves as a good first look. However, when higher quality reduction is needed, a manual version of the pipeline can be run that uses lines from an arc lamp observed before and after the object frames to perform wavelength calibration, as well as more recent flux standards that can improve atmospheric correction. \NEOx retrieves these data from the \LCO archive and automatically creates a reflectance spectrum with the most proximate Solar Analog spectrum observed with the same telescope as the object of interest.

\subsection{Development and Deployment Methodologies}
\label{sec:deploy}

One of the priorities for creating TOM (Target and Observation Management) systems like \NEOx is to allow them to be developed, adapted, and deployed by different groups for their own particular science interests and follow-up assets. Furthermore, scientific reproducibility is enhanced by having the full chain of software that was used to produce a scientific result available, along with the platforms the software operates on, which can be replicated by third parties via virtual machines \citep{Morris2017astrodocker}.

Mindful of the above, we have adopted the following philosophies:
\begin{itemize}
    \item the use of Python as the programming language,
    \item the use of version control (\texttt{git} and \textit{github.com}) to manage revisions to the software,
    \item the adoption of Test-Driven Development (TDD) methodologies to develop and maintain the software,
    \item the use of container technology to package and deploy the running software,
\end{itemize}{}

The NEOexchange platform is written in Python which has had a high take up for astronomy software and allows access to a large variety of astronomy-specific packages such as \texttt{astropy} \citep{AstroPy2013} and \texttt{astroquery} \citep{astroquery2019}. This is the most flexible choice at present for open access astronomy software.

Although the use of Test-Driven Development (TDD) in scientific software is currently small \citep{Nanthaamornphong2017test}, it can provide many benefits. The use of unit tests for the individual low-level functions and functional tests for the overall website operation helps in building more reliable and maintainable software over the long term. The combination of tests, along with the use of packaging and deployment technologies (described later), improves the reproducibility of scientific results by allowing others to reproduce them on a wide variety of platforms.  Given the use of the Python programming language for \NEOx, we decided to adopt \texttt{pytest} and \texttt{Selenium}\footnote{\url{https://www.selenium.dev/documentation/en/webdriver/}} for developing and running the unit and functional tests respectively. This builds on our previous efforts (e.g. \citealt{Lister2016}) but uses more modern web development and deployment frameworks such as Django and Docker to make a redeployable web service that can support many users and multiple projects that are using the \LCO Network. 

To allow the NEOx software to run in many more places and to decouple the code deployment from the underlying hardware and operating system, we make use of the Docker\footnote{\url{https://www.docker.com}} container platform to package and run the software. Docker has quickly emerged as the technology of choice for software containers.  Docker is an operating system level virtualization environment that uses software containers to provide isolation between applications (compared to virtual machines (VMs) which virtualize at the hardware level and require a guest operating system per VM). Through the use of a standardized file format (the \texttt{Dockerfile}) for describing and managing the setup of the containers, we can separate the sub-components of \NEOx (the database, web frontend, and tasks backend) into standardized containers for each component. 

This ability to be able to completely describe a software deployment environment with Docker has the potential to improve the reproducibility and the sharing of data analysis methods and techniques for the science and research community. This can allow other researchers the ability to reproduce and build on the original work (e.g. \citealt{Boettiger2015docker}) or to develop and distribute containerized versions of tools (e.g. \citealt{Nagler2015docker}) which can otherwise be hard to install and setup.

It is often the case that the research team developing the software does not have (or desire) direct control and management over the software environment where their software will be deployed. The deployment platform may be configured with different versions of the operating system, Python libraries, and modules than those that are required by their software, and yet are necessary for the stability and long-term support of other applications. This can present problems when attempting to update the version of these infrastructure components, as it is normally very difficult to isolate the different system components from each other, requiring all components to be updated at the same time.

With Docker, each application can be wrapped in a container configured with the specific version of operating system, Python interpreter, and Python packages and modules it needs to function properly. The common interface with the system is set at the container level, not at the operating system, programming language, or application server level.  In this container-based approach to deploying a service like \NEOx, the development process includes a container specifically designed for the service, with only the dependencies needed by the service.  The same container that is used during the development and testing of the \NEOx software on the developer's local machine can be deployed on a local computer cluster or in a private part of a cloud service, and even becomes part of the final project deliverable.  The final product is shipped and deployed as the container, with all of its dependencies already installed, rather than as an individual software component which requires a set of libraries and components that need to be installed along with it.  This not only simplifies the deployment of the final product, it also makes it more reproducible.

For \NEOx, we use Docker to build two sub-containers, one for fetching and compiling the \texttt{find\_orb}\footnote{\url{https://www.projectpluto.com/find_orb.htm}} orbit fitting code (which needs more modern compilers than our base CentOS\footnote{\url{https://www.centos.org}} 7 OS image possesses) and one for setting up the system and Python dependencies. The results are then incorporated into a third container which includes the necessary system tools (primarily \textsc{SExtractor}, \textsc{scamp}, and the CSS moving object code), the main \NEOx application code, and finally the \texttt{crontab} entries used for the object and data downloads (as described in Section~\ref{sec:ingest} and Section~\ref{sec:pipeline}) and other maintenance tasks.  The containers can be built and rebuilt using the Docker command-line tools and are also rebuilt automatically by the LCO-wide Jenkins\footnote{\url{https://jenkins.io}} automated build servers whenever commits are made to the \NEOx \texttt{github} repository. 

For deployment we make use of Amazon Elastic Kubernetes Service (EKS) to deploy our Docker containers to the Amazon Cloud and allow users to access the system. Kubernetes\footnote{\url{https://kubernetes.io}} is a container orchestration system for automating deployment and management of containerized applications. EKS makes use of Amazon's Elastic Compute Cloud (EC2) to provide the virtual servers for running the \NEOx application and the S3 storage system to store the files needed. We use Kubernetes to deploy several containers that setup and then run the \NEOx application. Two initialization containers are run at startup; one collects all of the static content (HTML, CSS, images, JavaScript) needed by the Django framework and webserver that runs the \NEOx application and the other is used to download the DE430 JPL ephemeris file \citep{Folkner2014de430} to the shared S3 storage for use by \texttt{find\_orb} during orbit fitting. Following the completed execution of these initialization containers, we deploy three other containers for the main \NEOx application. These are the \texttt{nginx} webserver, the backend \NEOx application itself, and the container for running the background and update tasks through the \texttt{crontab}. The Kubernetes system and the Amazon EKS handles constructing the needed services to direct web traffic from the internet to the \NEOx application and its webserver. It also handles redeploying the \NEOx containers should the application or one of the underlying virtual servers go down, or if an upgrade to a new version of our application is deployed. In the latter case, the new version is brought up in parallel and confirmed to be working before routing traffic over from the old version, at which point the older deployed version is removed. Kubernetes also allows us to run a developmental version of the \NEOx code in parallel, using a separated parallel database, without disturbing the production system. This allows us to build additional confidence that the system and any new features are working correctly, beyond the checks of our unit and functional tests, before making the new version live.

The use of a common, widely adopted programming language and web framework, coupled with modern software development methodologies, particularly Test Driven Development, has enabled us to develop and add many new features to the \NEOx codebase while ensuring that the code continues to work as expected. The adoption of containerizing technologies to package the software and all needed dependencies has also, after an initially steep learning curve, been greatly beneficial in that it has served to insulate the code and ourselves from many of the demands and trials of system administration. The more recent move to a more virtualized deployment strategy using cloud-based services, has similarly produced great medium and long-term benefits by allowing \NEOx to be kept continuously running during seamless and safe upgrades alongside parallel deployments for testing of new versions without impact to the production system.

\section{Use of the \LCO Network for NEO follow-up and characterization}
\label{sec:followup}

\subsection{Follow-up for NEO Surveys (PS1 \& 2, CSS, NEOWISE \& others)}
\label{sec:candfollowup}

One of the original motivations for building the \NEOx system (and its predecessor) was the large number of NEO candidates produced by the NEO sky surveys (thousands/year) requiring follow-up confirmation through astrometry and photometry. As described in Sections~\ref{sec:ingest} and \ref{sec:planning}, the \NEOx system retrieves new NEO candidates from the MPC, computes ephemerides, and can plan observations and automatically schedule them for follow-up on the robotic telescopes of the \LCONet.

The \NEOx homepage shows the candidates in need of follow-up, ranked according to the FOM equation (Equation~(\ref{eqn:FOM}) and Section~\ref{sec:planning}) along with recently confirmed NEOs. Individual candidates can be scheduled from here by following their associated links. There is also a command line user tool which can look at the current state of the candidate follow-up needs by querying the database and the state of the telescope network and then distributing candidates to available telescopes at a given time (planning for a future time is supported).
By default, this tool retrieves all of the NEOCP candidates that were ingested into the database within the last 5 days. In addition to older objects, we also filter out those that have not already been followed up and those that haven't gone more than 2.5 days without being observed. These cuts are necessary as the short $\sim30$\,minute discovery arc of typical NEOCP candidates means that their positional uncertainty can grow to many times our field of view after this time. All of these cuts can be customized by the user using command line options.

For each of the selected candidates, we compute the ephemeris and they are then filtered as follows:
\begin{itemize}
    \item $V$ magnitude in the range $19\leq V \leq22$,
    \item on-sky motion $\leq5\arcsec/\textrm{min}$,
    \item Moon-object separation $\geq30^\circ$,
    \item not already scheduled to be observed by \NEOx
\end{itemize}
Filtered objects are then split into four lists : first by declination (Objects with Dec. $<=+5^\circ$ go into the southern hemisphere list, objects with Dec.$>+5^\circ$ go into the northern hemisphere list),
 and then into a ``telescope class" list with $V\leq20.5$ objects going to the 0.4\,m telescopes and fainter to the 1\,m telescopes. This allows the targets to be directed to the telescope class that is optimal for their brightness.

Once the NEOCP candidates have been filtered and sorted, the currently dark subset of the \LCONet is checked for an operational status through an LCO-provided API. The targets are then scheduled on the appropriate site and telescope class. We operate by sending NEOCP candidates to a specific site, rather than using the more common mode of submitting to the whole \LCONet with a longer possible observing window, because we are interested in getting new measurements as soon as possible. By submitting to each site in turn as the Earth rotates we can optimize the time to a result and subsequent resubmits of the observation request will use the most current knowledge of the candidate's orbit at that time. We continue submitting requests for observations until either we get data or the object is confirmed by other observers and designated by the MPC.

After observation of the requested targets has occurred and the data has been processed as described in Section~\ref{sec:pipeline}, candidates that have been automatically detected by the moving object code are presented in \NEOx. After checking the validity of the observations by including them in a combined fit with the existing measurements (downloaded from \NEOx) using the \texttt{find\_orb} orbit fitter, the measurements are sent to the MPC. For targets which are not automatically identified or which are moving too fast to be identified in the individual frames and which require stacking on the object's motion, the data is downloaded and analyzed on a local workstation. Manually detected objects have their measurements checked for validity and then are reported to the MPC in the same way as described above.

\subsection{Rapid response to close-passing objects}
\label{sec:rapidresponse}

The combination of a large number of telescopes distributed around the world, coupled with the ability of the \LCO scheduler and the site software to re-plan, schedule and start execution of a new observing request within $\sim15$\,minutes, allows the \LCO network to respond rapidly to new targets of opportunity. These objects are primarily those NEO candidates that are determined to have a possibility of impact by the JPL SCOUT
early alert system. Since the particular candidate will already be in the \NEOx system from an earlier ingest of candidates (see Section~\ref{sec:ingest}), all that is necessary is to trigger a \textit{disruptive rapid-response} request. This involves setting a particular value of the observing mode in the observing request which tells the \LCO system that this request is of high enough priority that it can potentially disrupt and cancel an already running request at the telescope. This can be done through either the web frontend when scheduling an object or by adding a flag to the command line tool that was described in Section~\ref{sec:candfollowup}.

These rapid-response observations are considered first by the \LCO scheduler in a separate scheduling run and sent out to the sites, before the rest of the observing requests are scheduled, which reduces the latency. Once executed at the telescope, the resulting data is treated in the same way as that from other targets.

As briefly discussed  for NHATS and SCOUT targets in Section~\ref{sec:ingest}, there is a need for a better machine-readable format for requesting transient follow-up generally. Although the original SCOUT email alerts (which were machine generated and also semi-structured and machine readable to some extent) have now been supplemented by an API endpoint, the subsequent discussion about the nature of the object and validity of the alert and recommendations for continuing or halting follow-up are all conducted through free-form email conversations. 

This is also the situation in a number of other of time-domain and transient astronomy fields and illustrates the need to take the ``next step" to allow better, faster, and more automated responses to, and prioritization of, new transients. This extends the potential for automation and faster response further down the chain from the producers of alerts and the broker/triage stage to follow-up and characterization resources, as has already been done by many surveys e.g. ZTF \citep{Patterson2019}.

\subsection{Follow-up of radar-targeted NEOs and Potential Mission Destinations}
\label{sec:characterization}

Radar observations are a very powerful tool that are used to spatially resolve NEO targets, determine binary fraction and improve orbits, the last of which helps us to avoid loosing the target and improves impact risk assessment (e.g. \citealt{Ostro2007}). The \LCONEO allows rapid response astrometry that makes pointing and imaging by radar assets (such as Goldstone and Arecibo) possible for targets that have only recently been discovered. Potential mission destinations such as NEO Human Space Flight Accessible Targets Study (NHATS) targets are often small in size with a limited visibility window of days to a few weeks. The \LCONEO can quickly respond for characterization efforts for these objects. The network can also obtain colors and photometric light curves which allows the determination of rotation rates, pole directions, spectral classes, and shapes as well as perform robotic spectroscopy to determine taxonomic classes. The majority of our time allocation for the project is now spent on characterization efforts for these types of targets.

\begin{figure}
\centering
\begin{subfigure}%{.5\textwidth}
  \centering
  \includegraphics[width=.49\textwidth]{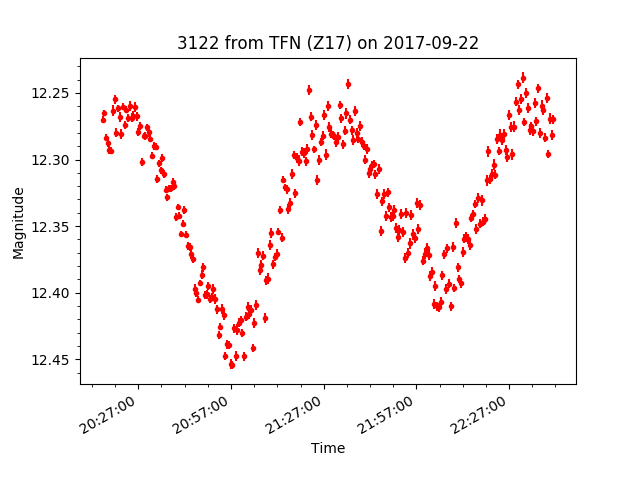}
%  \caption{A subfigure}
  \label{fig:lc_florence}
\end{subfigure}
\begin{subfigure}%{.5\textwidth}
  \centering
  \includegraphics[width=.49\textwidth]{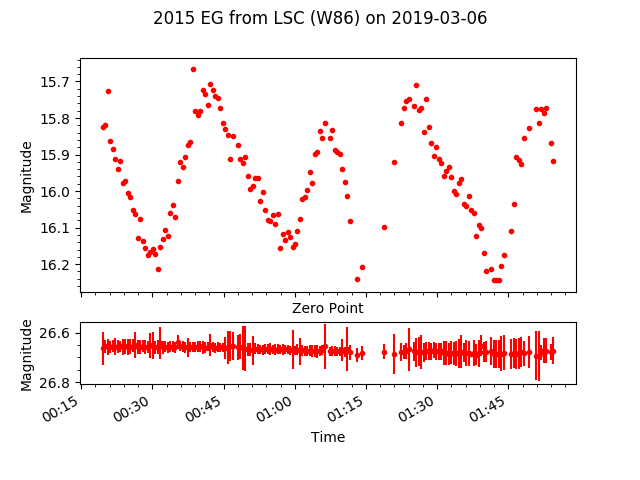}
%  \caption{A subfigure}
  \label{fig:lc_2015EG}
\end{subfigure}
\caption{Example NEO light curves obtained with \LCONEO: Light curve from the 0.4-m telescope in Tenerife (MPC site/telescope code Z17) of NEO (3122) Florence (left) which was observed with the Goldstone radar in 2017 August--September and was discovered to be a rare triple system and a light curve  of NEO 2015 EG from the \LCO 1-m in Chile (MPC site/telescope code W86) on 2019 March 6 (right) which was also observed that month by the Arecibo planetary radar. The right figure shows an example of the automatically produced plot products, with the light curve  plotted on top with the pipeline-determined per-frame zeropoint underneath.}
\label{fig:example_lc}
\end{figure}

In addition to the ingest of the characterization targets discussed in Section~\ref{sec:ingest}, we also ingest any known physical parameters such as a light curve period and amplitude, pole orientation, and taxonomic class if known. This is performed via a query and parsing of the JPL Small Body Database and storage of the results. In the absence of a period, we perform an initial observation block of 1\,hr in the PanSTARRS-w filter for maximum throughput, with the exposure time automatically calculated based on the object's speed as described in Section~\ref{sec:planning}. Following the results of these initial observations, further follow-up falls into one of three categories:
\begin{enumerate}
    \item The block covered at least $\sim1.5\times$ the period: period is determined and announced to community. Move onto taxonomic spectroscopy (for $V\lesssim 17$) or colors (for $V\lesssim 20.5$).
    \item Initial block indicates period in the $1\ldots4$\,hr range: Longer block scheduled to capture the full period.
    \item No indication of a period: Cadence scheduled covering 2--3 days with observations every 20--60\,minutes.
\end{enumerate}{}
Examples of the follow-up photometry for period determination that were obtained for two radar-targeted NEOs are shown in Figure~\ref{fig:example_lc}. (3122) Florence is a large ${\sim}4.2$\,km (using the MPC $H_V=14.1$ value and the NEOWISE albedo of 0.231; \citealt{Mainzer2011}) Earth-approaching Amor NEO which made a close (${\sim} 0.047$\,au) approach to the Earth in 2017 September; this was the closest approach until the 2057 apparition, making it an important opportunity to study this potentially very damaging large NEO. It was also the target of a citizen science Asteroid Tracker campaign for that year (see Section~\ref{sec:citizensci}).

Observations were carried out with the southern 1-meter telescopes and northern 0.4-meter telescopes of the \LCONet prior to and following close approach respectively. One of the light curves from one of the northern 0.4-meter telescopes in Tenerife (which has a MPC site/telescope code of Z17) is shown in Figure~\ref{fig:lc_florence}. These data confirmed the 2.36\,hr period found in the 1996--1997 apparition by \cite{Pravec1998}. (3122) was subsequently shown to be a triple system with 2 small companions in radar images.

2015 EG was discovered by Catalina Sky Survey's Mount Lemmon Station in 2015 March and its Earth-crossing short period ($P{\sim}293$\;day) and eccentric ($e{\sim}0.36$) Aten orbit means it makes frequent close approaches to both Earth and Venus. It was targeted by the Arecibo planetary radar on the next return in 2019 March, five orbital periods later. A 1.7 \;hour observation block two days after closest approach with the \LCO 1-meter at Cerro Tololo (MPC site/telescope code W86) showed a complex light curve with two unequal maxima and minima (Figure~\ref{fig:lc_2015EG}). The existence of distinct maxima and minima in the light curve  suggests an elongated shape but the fact that the two maxima and two minima are themselves unequal, shows that the situation is more complicated. This would need a more thorough light curve  inversion, ideally incorporating the radar data in a combined solution. We determine a period of ${\sim}0.717$\;hours based on the two maxima, in contrast to the previous determination of 1.29\;hours of \cite{Thirouin2016}, which was an estimate based on incomplete phase coverage. The radar ranging also secured a provisional weak detection of the Yarkovsky acceleration (see next section) on this object, making the Yarkovsky acceleration a more secure $2.4\sigma$ detection.

\subsubsection{Yarkovsky measurements}
\label{sec:yarkovsky}

The semi-major axis drift some asteroids undergo due to the non-gravitational Yarkovsky effect, caused by the unequal absorption and re-radiation of thermal energy on a rotating asteroid, can have a large influence on determining impact probabilities for asteroids that pass close to the Earth (\citealt{Giorgini2002, Giorgini2008}; \citealt{Farnocchia2013yarko}; \citealt{Farnocchia2014}; \citealt{Chesley2014}; \citealt{Spoto2014}; \citealt{Vokrouhlicky2015}). Even a small drift in semimajor axis can make the difference between a near-miss and an impact for close-passing asteroids when calculating long-term impact probabilities, especially when planetary close encounters perturb the orbit. 

The various command-line tools available through \NEOx (described in Section~\ref{sec:db}) were used to calculate the observability, brightness, and rate of motion of 36 asteroids determined to yield a high likelihood of a detection of the non-gravitational Yarkovsky effect \citep{Greenstreet2019}. The targets were ingested into the \NEOx database, which allowed for ephemeris calculation, planning, and scheduling of observations on the \LCONet.

\subsubsection{Robotic spectroscopic characterization}

Through the robotic FLOYDS spectrographs on the two 2-m telescopes on Maui, HI and Siding Spring, Australia (see Figure~\ref{fig:networkmap}), we can perform rapid-response low resolution spectroscopy of NEOs and other asteroids. After scheduling of spectroscopic observations (described in Section~\ref{sec:planning}) and data reduction (described in Section~\ref{sec:pipeline}), the data products for the target and the solar analog are extracted and made available through the \NEOx website from the requested observation block. This also allows display and download of the acquisition and guide movies made from the autoguider frames taken during the observations.

\begin{figure}
\centering
\begin{subfigure}%{.5\textwidth}
  \centering
  \includegraphics[width=.49\textwidth]{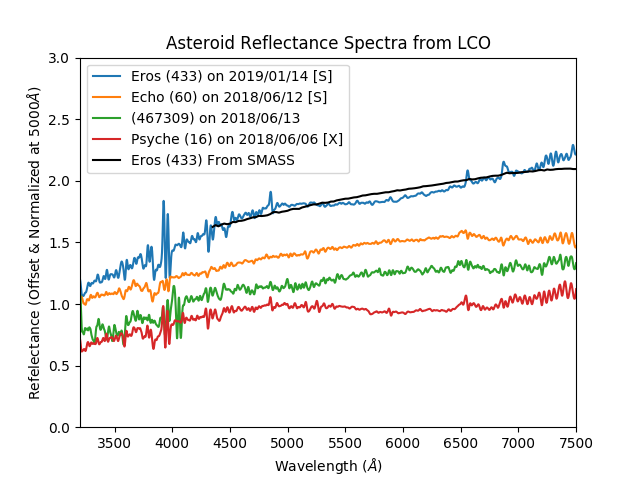}
%  \caption{A subfigure}
  \label{fig:eros_FLOYDS}
\end{subfigure}
\begin{subfigure}%{.5\textwidth}
  \centering
  \includegraphics[width=.49\textwidth]{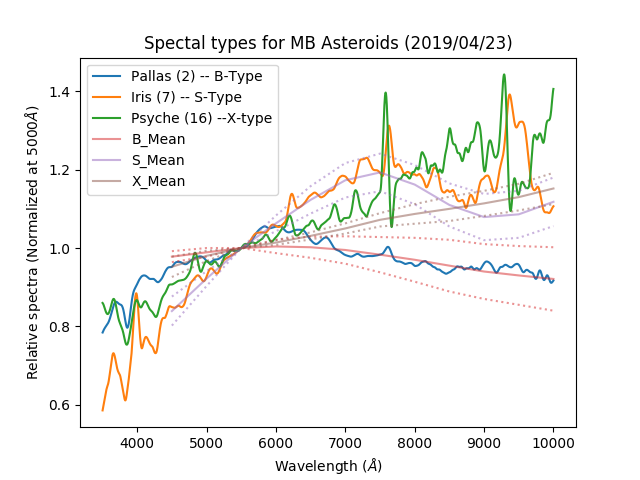}
%  \caption{A subfigure}
  \label{fig:MBAs_FLOYDS}
\end{subfigure}
\caption{Examples of optical reflectance spectra of NEO and Main Belt asteroids obtained with the FLOYDS spectrographs. (left) Normalized reflectance of NEO (433) Eros (blue) along with its optical/NIR template spectrum from SMASS (black line). Also shown are some other examples of S- and X-type asteroids. (right) Spectra of (2) Pallas, (7) Iris and (16) Psyche, along with the boundaries of the mean B-, S-, and X-type taxonomic classes from SMASS (dotted lines).}
\label{fig:example_spectra}
\end{figure}

The \NEOx site also allows a normalized reflectance spectrum to be automatically generated with the nearest-in-time suitable solar analog automatically chosen, plotted, and compared to taxonomic class templates from the SMASS library \cite{DeMeo2009, SMASSlibrary}. The plotting tool in the website permits users to interactively pan and zoom the spectra, and allows customized zooms of the plotted spectra to be saved to the user's computer. We plan to add the ability to reports wavelength and reflectance values at each datapoint and display known important lines from line lists such as atmospheric $\textrm{O}_2, \textrm{O}_3$, \&  $\textrm{H}_2\textrm{O}$ and mineral spectra. Some example NEO and Main Belt Asteroid reflectance spectra taken during the early commissioning phase of FLOYDS for moving objects are shown in Figure~\ref{fig:example_spectra}.

\subsubsection{Campaign on 1999 KW4}
\label{sec:kw4}

We participated in both the first NASA Planetary Defense Coordination Office (PDCO) Planetary Defense exercise in 2017 October on 2012 TC4 \citep{Reddy2019tc4} and the second exercise in 2019 May on NEO (66391) 1999 KW4. Due to the approach geometry, with the NEO approaching from low in the Southern sky, LCO's Faulkes Telescope South (FTS) was one of the few telescopes available to observe it before close approach. In preparation for this event, we worked extensively with the \LCO software team to improve the acquisition and guiding performance of the FLOYDS low resolution spectrographs, for fast moving objects ($>10\arcsec/\text{min}$), with a particular view to enhancing the capabilities to obtain spectra in support of the PDCO exercise on 1999 KW4.

\begin{figure}
\centering
\begin{subfigure}%{.5\textwidth}
  \centering
  \includegraphics[width=.49\textwidth]{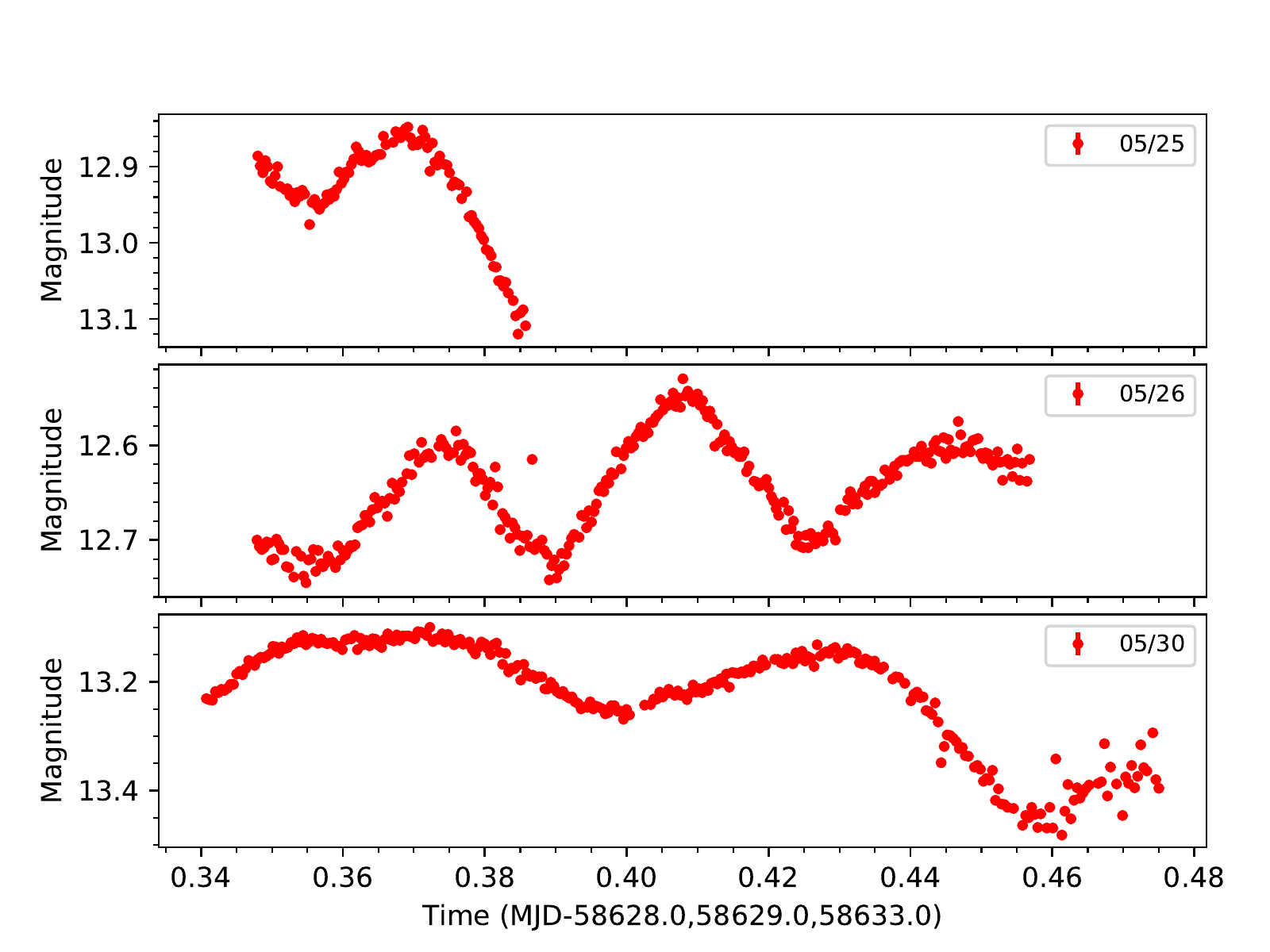}
%  \caption{A subfigure}
  \label{fig:kw4_lightcurves}
\end{subfigure}
\begin{subfigure}%{.5\textwidth}
  \centering
  \includegraphics[width=.49\textwidth]{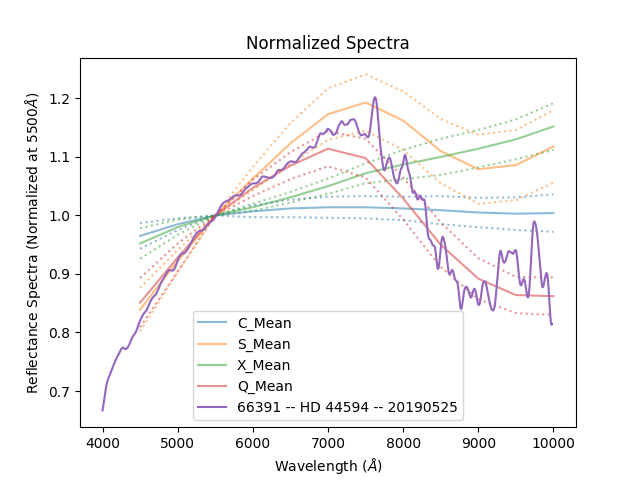}
%  \caption{A subfigure}
  \label{fig:kw4_spectra}
\end{subfigure}
\caption{Results from the \LCONEO on PDCO target (66391) 1999 KW4 (left) Light curves obtained with the 1m telescopes at Siding Spring Observatory, Australia on 2019 May 25, 26 and 30. The integer part of the MJD has been subtracted from the time of each datapoint and the remaining decimal day is used for $x$/time axis. A small number of outlier points have been clipped from each light curve where the target passed close to a field star or thin cloud intervened. (right)  Normalized reflectance of NEO (66391) 1999 KW4 (purple line) along with the boundaries of the mean C-, S-, X- and Q-type taxonomic classes from SMASS (dotted lines).}
\label{fig:kw4_figures}
\end{figure}

We were the only people able to get spectra before close approach and were able to get the data reduced and analyzed within a few hours of acquisition and circulated to the campaign participants approximately 17 hours before close approach. The FTS spectra showed that the taxonomic class of the NEO was also more like the rarer Q-type rather than the S-type previously measured (see Figure~\ref{fig:kw4_spectra}). We were also able to obtain a few light curves from several sites on the 1-meter network but extensive bad weather prevented us from gathering enough consecutive data from the southern sites of the \LCONet to provide full coverage of the 17.4 hour orbital period that would enable the full modelling of the system.

\begin{table}
\caption{Table of observations for (66391) 1999 KW4 with \LCO}
\begin{tabular}{cccccc}
\hline \hline
Block start and end & \LCO  & Telescope & MPC\footnotemark  & Observation & Num. \\
(UTC)  & Code  & Class & Site Code & Type & Exposures \\
\hline
%2018-05-15 08:10  $\rightarrow$ 08:54 & lsc & 0m4 & W89 & Opt. imaging & 295 \\
%2018-05-16 01:37 $\rightarrow$ 02:55 & cpt & 0m4 & L09 & Opt. imaging & 301 \\
%2018-05-17 00:45 $\rightarrow$  02:31 & cpt & 0m4 & L09 & Opt. imaging & 435 \\
2019-05-25 08:20 $\rightarrow$ 09:15 & coj & 1m0 & Q64 & Imaging & 111 \\
2019-05-25 09:11 $\rightarrow$ 09:16 & coj & 2m0 & E10 & Spectra & 1 \\
2019-05-25 09:25 $\rightarrow$ 09:30 & coj & 2m0 & E10 & Spectra & 1 \\
2019-05-26 08:20 $\rightarrow$ 10:57 & coj & 1m0 & Q64 & Imaging & 315 \\
2019-05-27 10:24 $\rightarrow$ 10:30 & coj & 2m0 & E10 & Spectra & 1 \\
2019-05-28 21:10 $\rightarrow$ 22:11 & tfn & 0m4 & Z21 & Imaging & 194 \\
2019-05-29 07:33 $\rightarrow$ 07:41 & ogg & 2m0 & F65 & Spectra & 1 \\
2019-05-29 22:00 $\rightarrow$ 23:14 & tfn & 0m4 & Z17 & Imaging & 217 \\
2019-05-30 08:10 $\rightarrow$ 11:28 & coj & 1m0 & Q63 & Imaging & 323 \\
2019-05-30 21:05 $\rightarrow$ 23:55 & tfn & 0m4 & Z21 & Imaging & 430 \\
2019-05-31 07:49 $\rightarrow$ 08:08 & ogg & 2m0 & F65 & Spectra & 1 \\
2019-06-03 16:36 $\rightarrow$ 19:26 & cpt & 0m4 & L09 & Imaging & 349 \\
\hline
\end{tabular}
\LCO Site Codes: coj: Siding Spring Observatory, Australia, tfn: Tenerife, Canary Islands, ogg: Maui, HI, cpt: South African Astronomical Observatory, South Africa
\end{table}
\footnotetext{\url{https://minorplanetcenter.net/iau/lists/ObsCodesF.html}}

During the preparation for the 1999 KW4 campaign, we developed code to construct on-demand long-term status plots for any object in the \NEOx database to show sky position, distance of the object from the Earth/Sun, magnitude and positional uncertainty. These will allow easier visualization and targeting of observing campaigns.

\subsection{Citizen Science project}
\label{sec:citizensci}

We used the software developed for this program to create Asteroid Tracker\footnote{https://asteroidtracker.lco.global}, with the aim of increasing public awareness of asteroid phenomena, timed with Asteroid Day 2016. This virtual event resulted in an audience of over 900 people signing up to each request observations on the \LCONet of two NEO radar targets we had selected for Asteroid Day 2016. This  provided a greatly simplified and streamlined version of the \NEOx portal for the general public allowing them to request a small number of observations on the network, using a single button, which would be combined together into a full-length sequence. 

Asteroid Tracker was enhanced after that event and was used for a larger scale event during Asteroid Day 2017 to target another NEO radar target (3122 Florence) as also described in Section~\ref{sec:characterization}. It was also used in a limited capacity to provide additional observations of (66391) 1999 KW4 during the planetary defense exercise described in Section~\ref{sec:kw4}. It has since been used as an ancillary material during the BBC/ABC TV series Stargazing Live in 2017.

The underlying code has since been refactored to use a custom TOM Toolkit build as the back-end, target and observation template source. This updated Asteroid Tracker front-end is a static website composed only of HTML, CSS and Javascript (unlike \NEOx which is requires a Django engine and database). The communication between front-end and back-end, is entirely asynchronously handled through APIs. The aim of this separation of back-end and front-end, was to provide multiple customised sites, with limited user options for citizen science projects, all back-ended by a single TOM.

We also created a project on the Zooniverse web platform\footnote{\url{ https://www.zooniverse.org }}, called \emph{Agent NEO}, using the Project Builder interface. This was a citizen science project to identify and report NEO candidates in \LCO data as described in Section~\ref{sec:candfollowup}. We used the Moving Target Detection algorithm to identify possible new NEOs in each follow up frame, as well as the target being followed up. For each candidate we created a cropped thumbnail image around the target, and uploaded the sequence of images. The Zooniverse users would step through the each candidate sequence and classified as either a genuine target or artifact. Each image sequence was given a unique identifier and stored in \NEOx, so the reported statistics from the Zooniverse classifications could be ingested back into \NEOx.

This was given to the 10,000 Zooniverse beta testing users in May 2017. We launched \emph{Agent NEO} on 2017 June 1 to tie in with Asteroid Day 2017 (30 June) and operated it for two months during the summer of 2017. A rapid cycle of redesign was necessary when creating \emph{Agent NEO} to match the \NEOx workflow to that of the Zooniverse Project Builder (chiefly the handling of animated image streams). Supporting such a large number of volunteers on the Zooniverse required considerably more person-effort than initially considered. We also encountered issues with the timeliness of the data flow through the \LCO and \NEOx pipelines and with large numbers of detector/flatfielding artifacts in the \LCO Sinistro cameras on the 1-meter telescopes. These artifacts mimicked the moving NEO candidates, which often have uncertain positions, making identification and recovery of the correct moving object NEO candidate difficult. This problem also affects the main \LCONEO program for candidates (Section~\ref{sec:candfollowup}. We had hoped that the situation would improve in 2018--2019 with the deployment of new Archon CCD controllers for the Sinistro cameras \citep{Harbeck2020} but this proved overly optimistic. This has been part of the reason for the change in focus of the \LCONEO from candidate confirmation towards NEO characterization. 

\section{Discussion and Future Work}
\label{sec:results}

We have described the design, implementation and operation of \NEOx, an online portal and TOM (Target and Observation Management) system for NEO and other small body science. We have also described the use of \NEOx to operate the \LCONEO conduct a program of follow-up to conduct a program of NEO candidate confirmation initially, with a transition into more detailed characterization of radar and other high value NEO targets, as usage and competition for time on the \LCONet increased.

Follow-up of NEO candidates with the \LCO telescopes over the July 2014 --July 2018 time period for which complete statistics are available from the MPC\footnote{\url{https://www.minorplanetcenter.net/mpc/obs_stats_2014_2018}} shows that 39300 measurements were reported. Of these, almost 9000 measurements were of nearly 1300 objects that were confirmed to be NEOs from Catalina, PS1, NEOWISE and other surveys. Although part of this period (before 2015 July) of follow-up of NEO candidates predates NEOexchange so we cannot search our database for these statistics (and the MPC database is not externally queryable in this way). Examining the subset of 468 reported confirmed NEOs in the \NEOx database and observed and reported to the MPC by \LCONet in 2015 July to 2018 July, shows that 310 of 468 ($\sim66$\%) had $\textrm{Dec}<0$ at discovery, showing the value of the additional resources of the \LCONet in the southern hemisphere.
 In the process of the follow-up, we have incidentally discovered over 100 new Main Belt objects and recovered objects as faint as $R\sim23$ showing the potential of the 1-meter network for NEO follow-up. 

We will be continuing to develop NEOexchange to allow use by other users (professional and citizen scientists), observation campaign planning and online data analysis and reporting. Some intended areas of improvement have already been discussed but they are also collected below:
\begin{itemize}
    \item additional interactive tools to allow users to perform period finding of light curves  and taxonomic determinations from asteroid spectra,
    \item automated stacking of frames for faster moving objects and detection of the asteroids in the resulting stacked images,
    \item investigation of the effect of 2nd-order astrometry effects such as DCR, proper motions and residual optical distortions (see Section~\ref{sec:pipeline}
    \item completion of the development of the generalized Exposure Time Calculator (ETC) framework for representing the atmosphere above a site, telescope and instruments and allowing more accurate prediction of needed exposure time to reach a particular SNR (Section~\ref{sec:planning}),
    \item use of the above ETC framework to allow modelling of instruments on larger telescopes such as the Goodman optical spectrograph on the SOAR 4.1-meter and the SCORPIO instrument (optical-NIR imager/spectrograph) on the Gemini South 8-m telescope as part of the Astronomical Events Observatory Network (AEON\footnote{\url{https://lco.global/aeon}}). 
    \item additional tools for coordination of characterization efforts such as aggregation and display of planned radar targets, citizen-science campaigns with Asteroid Tracker, follow-up observations from LCO, AEON other facilities and status of these efforts.
\end{itemize}

AEON aims to bring together a network of observatories to allow time-domain follow-up and science to be performed in a rapid, efficient and homogeneous way. It adds the SOAR 4.1-m telescope and Goodman spectrograph (optical) and the Gemini South 8-m telescope and likely the SCORPIO instrument (optical-NIR imager/spectrograph) once commissioned to the existing LCO telescope network of 0.4, 1, and 2-m telescopes, creating a more powerful ``virtual facility”. This ``virtual facility” becomes a powerful tool for the study of all kinds of solar system objects and transient behavior and  subsequent evolution with access to telescope resources from 0.4--8-m with optical and NIR imaging and low resolution spectroscopy instruments.

The production of transient alerts from current and future sky surveys, the integration of alert streams from alert brokers such as ANTARES \citep{Saha2014, Narayan2018}, with the target coordination and follow-up capabilities provided by TOMs such as \NEOx, and rapid response telescopes and data processing pipelines allows “closed loop” small body science to be achieved on exciting short timescale dynamic phenomena across the Solar System. It will also allow the building of a scaleable system that can perform classification and characterization observations on the increasing number of NEOs being discovered by current and future sky surveys during their often brief window of observability.

\section*{Acknowledgements}
Funding from the NASA NEOO program through grants NNX14AM98G and 80NSSC18K0848 to \LCOfull is acknowledged. We are very grateful to all of the \LCO Engineering, IT, and Software teams for designing and building the hardware and software that makes the \LCONEO possible. In particular we would like to thank Steve Foale for the improvements to the guider software to make the FLOYDS robotic spectroscopy of moving objects possible and Ira Snyder for the development of, and very valuable assistance with, Kubernetes and the \LCO Amazon Cloud infrastructure.  This research has made use of data and/or services provided by the International Astronomical Union's Minor Planet Center and the VizieR catalogue access tool, CDS, Strasbourg, France (DOI: 10.26093/cds/vizier). This work makes use of observations from the Las Cumbres Observatory global telescope network.

S. Greenstreet acknowledges support from the Asteroid Institute, a program of B612, 20 Sunnyside Ave, Suite 427,
Mill Valley, CA 94941. Major funding for the Asteroid Institute was generously provided by the W.K. Bowes Jr.
Foundation and Steve Jurvetson. Research support is also provided from Founding and Asteroid Circle members K.
Algeri-Wong, B. Anders, R. Armstrong, G. Baehr, The Barringer Crater Company, B. Burton, D. Carlson, S. Cerf,
V. Cerf, Y. Chapman, J. Chervenak, D. Corrigan, E. Corrigan, A. Denton, E. Dyson, A. Eustace, S. Galitsky, L. \& A.
Fritz, E. Gillum, L. Girand, Glaser Progress Foundation, D. Glasgow, A. Gleckler, J. Grimm, S. Grimm, G. Gruener,
V. K. Hsu \& Sons Foundation Ltd., J. Huang, J. D. Jameson, J. Jameson, M. Jonsson Family Foundation, D. Kaiser,
K. Kelley, S. Krausz, V. Lašas, J. Leszczenski, D. Liddle, S. Mak, G.McAdoo, S. McGregor, J. Mercer, M. Mullenweg,
D. Murphy, P. Norvig, S. Pishevar, R. Quindlen, N. Ramsey, P. Rawls Family Fund, R. Rothrock, E. Sahakian, R.
Schweickart, A. Slater, Tito’s Handmade Vodka, T. Trueman, F. B. Vaughn, R. C. Vaughn, B. Wheeler, Y. Wong, M.
Wyndowe, and nine anonymous donors.

S. Greenstreet acknowledges the support from the University of Washington College of Arts and Sciences, Department of Astronomy, and the DIRAC Institute. The DIRAC Institute is supported through generous gifts from the Charles and Lisa Simonyi Fund for Arts and Sciences and the Washington Research Foundation.

\section*{References}

\bibliography{lco_solsys}{}

\end{document}